\documentclass[twocolumn,aps,a4style]{revtex4}
\usepackage{epsfig}
\usepackage{amsmath}
\usepackage{amssymb}
\usepackage{amsthm}
\usepackage{graphics}
\usepackage{epsfig}

\newcommand{\beq}{\begin{equation}}
\newcommand{\eeq}{\end{equation}}
\newcommand{\deltap}{\eta}
\newcommand{\etap}{a_1}
\newcommand{\gammap}{a_2}
\newcommand{\lambdap}{a_3}
\newcommand{\Omegap}{\varpi}
\newcommand{\WW}{{\cal Z}}

\begin{document}

\title{Vacuum polarization in asymptotically anti-de Sitter black hole geometries}

\author{Antonino Flachi}
\email{flachi@yukawa.kyoto-u.ac.jp} 
\address{Yukawa Institute for Theoretical Physics, Kyoto University, Japan}
\author{Takahiro Tanaka}
\email{tama@scphys.kyoto-u.ac.jp} 
\address{Department of Physics, Kyoto University, Japan}
\preprint{YITP-07-87}
\preprint{KUNS-2132} 

\begin{abstract}
We study the polarization of the vacuum for a scalar field, $\langle
\phi^2 \rangle$, on a asymptotically anti-de Sitter 
black hole geometry. The
method we follow uses the WKB analytic expansion and point-splitting
regularization, similarly to previous calculations in the
asymptotically flat case. Following standard procedures, we write the
Green function,  regularize the initial divergent expression by
point-splitting, renormalize it by subtracting geometrical counter-terms, 
and take the coincidence limit in the end. 
After explicitly demonstrating  the cancellation
of the divergences and the regularity of the Green function, we express
the result as a sum of two parts. One is calculated analytically and
the result expressed in terms of some generalized zeta-functions, which
appear in the computation of functional determinants of Laplacians on
Riemann spheres. 
We also describe some systematic methods to evaluate these functions numerically. 
Interestingly, the WKB approximation naturally organizes 
$\langle \phi^2 \rangle$ as a series in such zeta-functions. 
We demonstrate this explicitly up to next-to-leading order in the WKB expansion.
The other term represents the `remainder' of the WKB approximation and
depends on the difference between an exact (numerical) expression and
its WKB counterpart. This has to be dealt with by means of numerical
approximation. The general results are specialized to the case of
Schwarzschild-anti-de Sitter black hole geometries. The method is
efficient enough to solve the semi-classical Einstein's equations 
taking into account the back-reaction
from quantum fields on asymptotically anti-de Sitter black holes.
\end{abstract}
\maketitle
%
%

\section{Introduction}
Classically, matter influences gravity via its stress-energy tensor,
$T_{\mu\nu}$, that appears as a source term in Einstein's equations. It
seems natural to expect that, if quantum fluctuations of matter
fields and the curvature are sufficiently small, this picture may remain
valid also in the semi-classical theory, where quantum fields propagate
on a given classical geometry. In this case, the quantum field would
couple to gravity via its stress-energy tensor, and its `back-reaction'
effects on the background spacetime would be described by the
{\it semi-classical} Einstein's equations:
$$
G_{\mu\nu} = 8\pi \langle T_{\mu\nu} \rangle~.
$$
Even though the precise range of validity of the above equations cannot
be quantified without knowledge of the full quantum gravity, one may
reasonably expect that, as long as back-reaction effects are locally
small, semi-classical Einstein's equations may still provide an adequate
description even if they lead to large cumulative effects, as in black
hole evaporation.

In general, calculating semi-classical back-reaction effects is not an
easy task, and it presents itself with a series of major
difficulties. Ref.~\cite{wald_book} lists three of them. The first is
related to the fact that in curved spacetime no renormalization
prescription can be uniquely given, and this generates an ambiguity in
the expectation value of the stress-energy tensor. Obviously, the
semi-classical approximation cannot resolve this issue. The second
problem is related to the fact that back-reaction effects introduce
higher-derivative terms into the classical equations, possibly
introducing spurious solutions. Therefore new criteria, beyond the
semi-classical approximation, have to be invoked to select the physically
relevant ones.
A third problem stressed in \cite{wald_book} is that: `It is very
difficult to compute $\langle T_{\mu\nu} \rangle$'. Even for highly
symmetric space-times, or for particularly simple vacuum states, it is a
formidable task to obtain $\langle T_{\mu\nu} \rangle$ even by numerical
approximation.

The problem of quantum back-reaction is particularly interesting in the
case of black holes, and, more generically, in spacetimes with horizons,
because particle creation may lead to significant effects that may cause
the evaporation of black holes. For this reason, since the original
discovery of Hawking \cite{hawk}, the study of quantum effects from
matter fields on black hole geometries has been the subject of much
attention.

What is really necessary for studying the effect of the back reaction is
the full averaged energy-momentum tensor. However, $\langle
\phi^2\rangle$, which we will term for convenience {\it vacuum
polarization}, already conveys much physical information, with the bonus
of a slightly less cumbersome computation.
Candelas was the first to calculate the polarization of the vacuum for
massless, minimally coupled scalar fields on a Schwarzschild geometry
outside the horizon, and obtained a renormalized expression for $\langle
\phi^2\rangle$ using the point-splitting method \cite{candelas}. The
initial work of Ref.~\cite{candelas} has been extended in various
directions. Further analysis of the renormalized vacuum polarization has
been done in \cite{candelas2} and the expectation value of the
energy-momentum tensor for the Hartle-Hawking state in the case of a
conformally coupled scalar field on the same geometry has been obtained
in \cite{candelas3,howard}. Other results were reported in
\cite{jensen1}, where the expression for the vacuum polarization has
been extended inside the black hole horizon. In Ref.~\cite{fawcett}
Fawcett calculated numerically the energy-momentum tensor for conformal
scalar fields and compared those results with Candelas' ones
\cite{candelas} and with Page's analytical approximation \cite{page,bpo}. 
Jensen and Ottewill have extended the above computation of the energy-momentum tensor to the electromagnetic field \cite{jo}.

Since these initial works, the study of black hole radiance kept, and
is still keeping, the interest in the problem alive. Novel results have
been obtained and extended the original ones to different background
geometries and higher spin fields.
Anderson considered the case of a massive field with general coupling and
calculated the vacuum polarization on a Schwarzschild background
\cite{anderson1} and on an asymptotically-flat, static, spherically
symmetric spacetime \cite{anderson2}, both in the case of Hartle-Hawking
vacuum. These results have then been extended and the full
energy-momentum tensor has been evaluated in \cite{anderson3}. Based on
the methods described in Ref.~\cite{anderson1,anderson2,anderson3},
Sushkov developed an analytical approximation and obtained $\langle
\phi^2\rangle$ for Schwarzschild and wormholes spacetimes
\cite{sushkov}. The analysis of quantum effects has also been extended
to consider various aspects of quantum effects in rotating black hole
geometries in \cite{chrzanowski,elster,zannias,frolov} and recently further work
in the case of Reissner-Nordstrom \cite{mottola} and lukewarm black
holes has also been studied \cite{winstanley}.

More recently, renewed interest in the quantum back reaction on 
black hole geometries has arisen in the context of the AdS/CFT correspondence. 
The five-dimensional brane world model proposed by Randall and Sundrum 
with an infinite warped extra-dimension~\cite{RSII} has been noticed 
to be possibly equivalent to the four-dimensional Einstein gravity 
with CFT corrections~\cite{HawkingReall}. 
If we believe this correspondence, gravity in the brane world can be 
understood by inspecting the quantum back-reaction due to CFT. 
Applying the correspondence to black holes, a conjecture has been proposed: a brane localized black hole in Randall-Sundrum brane world model 
cannot stay static~\cite{Tanaka:2002rb,Emparan:2002px}.
In this context, the energy-momentum tensor due to 
a free scalar field on a four-dimensional Schwarzschild background in the Boulware 
state, which is compatible with the asymptotic flat condition, 
was computed, and it was confirmed that the quantum back-reaction 
diverges on the horizon~\cite{Fabbri:2005zn}.
Although there have been many 
attempts to examine whether this conjecture is correct or
not~\cite{Kanti:2001cj, Karasik:2003tx,
Dadhich:2000am, Casadio:2001jg, Chamblin:2000ra, Shiromizu:2000pg,Kudoh:2003xz,Kudoh:2004kf}, 
we have not reached any definite conclusion yet.    
The original conjecture is about asymptotically flat brane geometries. 
One of the present authors suggested that one may be able to shed new light on this issue 
by extending the setup to asymptotically AdS brane geometries~\cite{Tanaka:2007xm,Karch:2000ct}. 
What will happen on the CFT side has already been discussed by 
Hawking and Page many years ago~\cite{HawkingPage}. In this case 
it is expected that a static black hole solution exists 
even after taking into account the quantum back-reaction 
if the size of the black hole is sufficiently large.
However, such a self-consistent quantum black 
hole solution has not been realized numerically so far.  

The above reasons provided enough motivation for us to consider the
effects of quantum fields on asymptotically anti-de Sitter (AdS) black hole
geometries. We will not solve the full back-reaction  problem in this
notes, rather we will focus on the computation of the vacuum
polarization for massive, non-minimally coupled scalar fields, leaving
the details of the full energy-momentum tensor calculation and of the
effects of the quantum back-reaction to a forthcoming publication \cite{progress}.

The strategy presented in this paper is valid for a general spherically
symmetric, asymptotically AdS spacetime, but, for brevity, we focus our
attention to the case of Schwarzschild-anti-de Sitter (Sch-AdS) black
holes, described by the following form line element:
\beq
ds^2=-f(r)dt^2+f^{-1}(r)dr^2+r^2d\Omega^2~,
\label{metric}
\eeq
where
\beq
f(r) = 1-{1\over m_p^2}{2M\over r} + {r^2 k^2}~,
\label{schads}
\eeq
$m_p=G^{-1/2}$ is the Planck mass in natural units and $k^{-1}$ is the
AdS curvature length. 
The horizon is located at $r=r_h$, where $f(r_h)=0$. Customarily, the
time coordinate is complexified, $\tau=it$, rendering the metric
positive definite outside of the horizon. The apparent singularity at
the horizon is removed by regarding the $\tau$ coordinate as periodic
with period $\beta$:
$$
\beta=4\pi\left[{df\over dr}\right]^{-1}_{r=r_h}~,
$$
that, for Sch-AdS black holes, becomes
$$
\beta=4\pi{r_h\over 1+3k^2r_h^2}~.
$$
Analogously to the asymptotically flat cases, asymptotically AdS black holes have a characteristic temperature, $T=\beta^{-1}$, and an
intrinsic entropy equal to a quarter of the area of the event
horizon. However, unlike the asymptotically flat black holes, there
exists a minimum temperature which occurs when its size is of the order
of the characteristic radius of the AdS space. As a working
example, we will consider the above AdS black holes and the numerical
analysis will be specialized to this case.

The results we will present are technical and mainly follow the direct
procedure outlined by Candelas in the case of Schwarzschild black
holes. The Green function is expressed as a sum of the products of the normal modes, therefore the first step in the computation requires the homogeneous
solutions of the matter field equations to obtain a complete set of solutions. The choice of the normal modes selects a specific vacuum state. 
Although the wave equation is separable in the spherically symmetric case, 
it is not in general analytically soluble, and some numerical analysis is necessary.
However, to practically perform renormalization of the Green function, which
originally diverges quartically, we need some analytic approximation 
to the normal modes. For this purpose, the WKB approximation is useful. In the following we illustrate how to obtain the WKB approximants for the normal modes, 
highlighting the differences with the asymptotically flat case. 

After this initial part, we discuss the renormalization of the Green
function. To subtract the divergent geometrical counter-terms, we 
first need to regularize the divergent sum.  
For this purpose, an appropriate procedure is the point splitting
regularization, developed in general by Christensen
\cite{christensen}. Relaxing the regularization after subtracting 
the divergent pieces, we obtain a renormalized expression
for the coincidence limit of the Green function, which is formally
divided into two parts: one amenable to analytical computation and one
that requires numerical evaluation. The latter part can be arranged 
so as to get fast convergence in the mode sum. 

The point splitting method guarantees the regularity of the Green
function. Customarily the explicit proof of the regularity is not
illustrated. 
Despite of the apparent superfluousness, 
we explicitly perform this computation, which, aside of
being a non-trivial check of a complex algebraic calculation, 
will turn out to be useful to convince the reader of our systematic way of 
evaluating the renormalized vacuum polarization. 
In fact, the renormalized propagator is made of various
divergent parts, which, combined with each other, lead to a regular
expression. Recognizing the structure of the divergences for each of
these pieces allows us to perform a piece-by-piece renormalization. This
will allow us to handle the numerical computation in a more convenient
way. 
In fact, as we will see, the analytical part of the propagator can
be expressed in terms of certain regularized $\zeta-$functions, which
appear in the computation of determinants of Laplacians on Riemann
spheres~\cite{elizalde}. 
The conclusive computation of the vacuum polarization is illustrated in Sec.~\ref{sec6} where the numerical and analytical calculations are presented in detail. Our final remarks close the paper.

\section{Green function}
\label{sec3}
The Euclidean Green function for a scalar field satisfies the following equation:
$$
\left(\Box -m^2 -\xi R\right) G_E(X,X') = -\sqrt{g}\delta(X,X')~.
$$
Choosing as vacuum state the Hartle-Hawking one, it can be expressed as
\begin{eqnarray}
G_E(X,X')& = &{1\over \beta} \sum_n e^{2\pi in(\tau-\tau')/\beta} 
\cr
&&\times\sum_l{(2l+1)\over 4\pi }P_l(\cos \gamma) G_{nl}(r,r')~,
\label{gaux}
\end{eqnarray}
which is characterized by the regularity and the periodicity along the $\tau$ direction with period $\beta$. Here, 
$\cos \gamma = \cos \theta \cos \theta'+\sin \theta \sin \theta'\cos (\phi-\phi')$. 
Writing the delta function as
$$
\delta(\tau,\tau') = {1\over \beta} \sum_n e^{2\pi
in(\tau-\tau')/\beta}~, 
$$
one can find that the function $G_{nl}$ satisfies 
\begin{eqnarray}
\biggl\{{d\over dx}\left(x^2 f {d\over dx}\right)-l(l+1) 
-{x^2\omega_n^2\over f}\qquad\qquad &&
\cr 
-(\hat{m}^2+\xi \hat{R})x^2 \biggr\}G_{nl}(x,x')=-\delta(x-x')~.&&
\label{radeq}
\end{eqnarray}
In the above equation we have rescaled the radial coordinate, the mass
and the curvature as:
$$
x=kr~,
\qquad \hat{m}=m/k~,
\qquad 
\hat{R} = k^{-2}R~,
$$
and defined
\beq
\omega_n=\alpha n~,
\label{omegaenne}
\eeq
with
$$
\alpha={2\pi \over k\beta}~,
$$
which for Sch-AdS black holes takes the following form:
$$
\alpha= {1\over 2}\left({{1\over x_h}+3x_h}\right)~,
$$
where $x_h=kr_h$. 
The function $G_{nl}(x,x')$ can be written in the usual way as the
product of the two independent solutions of the homogeneous equation
associated with Eq.~(\ref{radeq}):
\begin{eqnarray}
\biggl\{{d\over dx}\left(x^2 f {d\over dx}\right)-l(l+1)
 -{x^2\omega_n^2\over f}\qquad\qquad&&
\cr
-(\hat{m}^2+\xi \hat{R})x^2 \biggr\}\varphi(x)=0~.&&
\label{varphieq}
\end{eqnarray}
Using tortoise coordinates,
$$
dx_* = dx/f~,
$$
the homogeneous equation can be written as
\beq
\left[{d^2\over dx^2_*}-\omega_n^2-\left({l(l+1)\over x^2}+ {f'\over
x}+\hat{m}^2+\xi \hat{R}\right)f\right](x\varphi)=0~,
\label{eqrw}
\eeq
where $'$ represents derivative with respect to $x$. It is instructive to examine the behavior of the solutions near the horizon and at infinity. In the near horizon region the solutions are determined by
\beq
\left[{d^2\over dx^2_*}-\omega_n^2\right](x\varphi)\sim 0~.
\label{hori}
\eeq
This leads, as in the asymptotically flat case, to exponential solutions, 
\beq
\varphi\sim e^{\pm \omega_n x_*}/x~,
\label{expsol}
\eeq
one of them being regular (at the horizon). At infinity (large-$x$) one finds
\beq
\left[{d^2\over dx^2}+{4\over x} {d\over dx}-{1\over
x^2}\left(\hat{m}^2+\xi \hat{R}\right)\right]\varphi\sim 0~,
\eeq
which, differently from the asymptotically flat case, admits solutions of the form
\beq
\varphi\sim x^{-{3\over 2} \pm {1\over 2}\sqrt{9+4(\hat{m}^2+\xi
\hat{R})}}~.
\label{-3}
\eeq
Going back to the original equation (\ref{radeq}), its solutions should be
chosen by specifying their asymptotic behavior so that they are regular
at infinity and on the horizon. We indicate as $p_l^n(r)$ the solution
regular on the horizon and as $q_l^n(r)$ the one regular at infinity. 
The WKB approximation of these solutions will be discussed 
in detail in the next section. Notice that, in the Schwarzschild case, 
Eq.~(\ref{radeq}) takes
the form of Heun equation, with two regular singular points and one
irregular singularity at infinity. The 4D Sch-AdS case is
different. Eq.~(\ref{radeq}) has five regular singular points, with
infinity being a regular singularity.

The $n=0$, conformally coupled case is somewhat special, because at the horizon the zero order WKB approximant vanishes in the limit $\xi=1/6$ and $m=0$. In the $n=0$ case, the behavior at the horizon can be easily understood by re-expressing the mode equation in terms of the the logarithmic derivative of $\varphi$ near the horizon. It takes the form
$$
x^2 f' {d \ln \varphi \over dx} - l(l+1)-2x^2 = 0~,
$$
from which it is easy to read off the behavior at $x=x_h$:
$$
\left[{\varphi'\over \varphi}\right]_{x_h} = {l(l+1)-2x_h^2\over x_h+3x_h^3}~.
$$

Standard procedure leads to the following expression for the Green function:
\begin{eqnarray*}
G_E(X,X')&=&{\alpha k^2\over 2\pi} \sum_n e^{in\alpha\epsilon} \sum_l{(2l+1)\over 4\pi }P_l(\cos \gamma)
\cr
&&\times  {p_l^n(x_<) q_l^n(x_>)\over 
x^2 f \left(q_l^n dp_l^{n}/dx-p_l^n(x) dq_l^{n}/dx\right)}~,
\end{eqnarray*}
where $\epsilon=k(\tau-\tau')$.
The above expression is yet formal, because we have not found any
explicit solution for the mode functions. In fact, as one can easily
guess, such a solution can only be found numerically or by using some
approximation method. 
%
%

\section{WKB solutions}
\label{sec4}
A convenient way to obtain an analytic expression for the 
solution of Eq.~(\ref{varphieq}) is to use the WKB approximation. 
As we will see, this approximation is suitable and sufficient to perform
all the renormalization procedure that we will need later.  

The WKB method can be implemented by writing the solution as
\beq
\varphi(x) = x^{-3/2} W^{-\deltap} e^{\pm\int_{x_h}^x {W(x') h(x')}dx'}~,
\label{wkb}
\eeq
with $\deltap>0$. The $+$ sign refers to the solution regular at the
horizon, which we have called $p_l^n(x)$ and the $-$ sign to
$q_l^n(x)$. The overall factor $x^{-3/2}$ is multiplied so that the
asymptotic form at $x\to\infty$ becomes compatible with the 
WKB ansatz: $\varphi\sim x^{-{3\over 2}} 
e^{\pm{1\over 2}\sqrt{9+4(\hat m^2+\xi\hat R)}\log x}$. 
Substituting the previous ansatz in the homogeneous equation
associated with (\ref{radeq}), one easily finds
\begin{eqnarray}
0&=&W^2 \pm \left(-{3 \over x h}+{h'\over h^2}+{(x^2f)'\over x^2f h}\right)W
\pm{1-2\deltap\over h}W'
\cr &&
+\deltap\left({3\over xh^2}-{(x^2f)' \over x^2f h^2}\right){W'\over
W}-{\deltap \over h^2}{W''\over W}\nonumber\\
&&+{\deltap(1+\deltap) \over h^2}{W^{'2}\over W^2}
+{15\over 4x^2 h^2}- {3(x^2f)'\over 2x^3f h^2}
-{l(l+1)\over x^2fh^2} 
\cr &&
-{\omega_n^2\over f^2h^2}+\hat{m}^2{f\over x^2}+\xi \hat{R}{f\over x^2}~.\nonumber
\end{eqnarray}
We choose the coefficient $\deltap$ and the function $h$ 
so as to cancel the term with ($\pm$)-signature. 
Hence, we have $\deltap =1/2$ and $h= x/f$, and 
the WKB equation simplifies to 
\begin{eqnarray}
W^2&=&\Omegap+ \sigma+ \etap {W'\over W}+\gammap{W^{'2}\over W^2}+ \lambdap{W''\over W}~,
\label{sei}
\end{eqnarray}
where
\begin{eqnarray} 
\Omegap &\equiv& \left[
\left(l+{1\over 2}\right)^2-{1\over 4}\right]{f\over x^4}+{\omega_n^2\over x^2}~,\nonumber\\
\sigma&\equiv& {3\over 2}{ff'\over x^3}- {3\over 4} {f^2\over
 x^4}+\hat{m}^2{f\over x^2}+\xi \hat{R}{f\over x^2}~, \nonumber
\end{eqnarray}
and we have defined 
\begin{eqnarray*}
\etap  \equiv{f\over 2x^2}\left(f'-{f\over x}\right)~,\quad
\gammap\equiv-{3\over 4}{f^2\over x^2}~,\quad
\lambdap\equiv{f^2\over 2x^2}~.
\label{defa}
\end{eqnarray*}
As it is well-known, the WKB method expresses the solution iteratively
as 
$$
W=W^{(0)}+W^{(1)}+W^{(2)}+\cdots~.
$$
The leading order term is 
$$
W^{(0)} = \sqrt{\Phi(l)}\equiv \sqrt{\Omegap+\sigma}~,
$$
the next-to-leading order correction is computed by adding the derivative terms
in (\ref{sei}) evaluated for $W= W^{(0)}$, and so on, iteratively, to
the desired order. In the following, everything is calculated up to
next-to-leading order, {\it i.e.} $W^{(1)}$.

To be more precise, we introduce the following truncated WKB
approximation of $1/W$:
\begin{eqnarray}
{1\over W_n(z)}\equiv {1\over \Phi(z)^{1/2}}-{\Psi(z)\over 4\Phi(z)^{3/2}}~,
\label{dsw1}
\end{eqnarray}
where the function $\Psi(z)$ conveys the next-to-leading order WKB corrections and is defined by 
\begin{equation}
\Psi \equiv a_1 {\Phi'\over \Phi}+ \tilde a_2{\Phi^{'2}\over \Phi^2}+ 
 a_3{\Phi''\over \Phi}~, 
\label{defPsi}
\end{equation}
where 
\begin{eqnarray}
\tilde a_2\equiv{\gammap-\lambdap\over 2 }=-{5\over 8}{f^2\over x^2}~.
\label{defalpha}
\end{eqnarray}
Going to higher order in the WKB expansion means adding higher order
terms obtained by reiterating the above procedure. 
The explicit WKB expansions are 
lengthy and we will not report them here, but they can easily be handled
by any symbolic algebra manipulation program. 

One can easily check the quality of the WKB approximation by comparing
it with the numerical solutions. It turns out that the numerical
solutions are well reproduced by the WKB approximation already for small 
values of $n$ and $l$, as it is easy to
argue. In fact, expanding $W$ to next-to-leading order, one can easily see that the remainder in $1/W$ is of
order $O(l^{-5},n^{-5})$. 
%
%

\section{Coincidence Limit} 
\label{sec5}
Our main goal is to compute the coincidence limit of the Green function
which will provide us the expression for $\langle \phi^2 \rangle$. Here
we follow the same procedure that has been used in the Schwarzschild
case. First take the partial coincidence limit, by setting $r=r'$ and $\Omega=\Omega'$. In this case the
expression for the Green function simplifies to
\beq
G_E(X,X')={k^2\alpha\over 8\pi^2} \sum_{n=0}^\infty e^{in\alpha\epsilon} \sum_{l=0}^\infty{(l+1/2) \over x^3 \tilde{W}_{n}(l)}~,
\label{ge2} 
\eeq
where 
$$
{1\over {\tilde W}_{n}(l)} \equiv
{ 2x\over f} 
\left({d\ln q_l^n(x)\over dx}
     -{d\ln p_l^n(x)\over dx} \right)^{-1}~.  
$$
In the above expression the coincidence limit cannot be taken in a
straightforward way because the sums over $l$ and $n$ are divergent. The
divergence related to the $l$ summation is not a serious one and can be
by-passed easily by noticing that the Green function is a finite and
well-defined object as long as $X\neq X'$, i.e. $\epsilon \ne 0$.  
The trick commonly used is to
add a multiple of $\delta(\tau-\tau')$ (and its derivatives if
necessary) which, as long as
$\tau\neq\tau'$, does not alter the result. Basically, we can add
terms of the form:
$$
{k^2\alpha\over 8\pi^2} \sum_n e^{in\alpha\epsilon}\sum_l R_l(x)
$$
where a function $R_l(x)$ independent of $n$ can be chosen freely. 
By looking at the asymptotic behavior of the solution for large $l$, 
which can be obtained from the WKB approximation, 
$$
\tilde W_{n}(l)\sim  
W_{n}(l) 
\sim (l+1/2)f^{1/2} x^{-2}+O\left((l+1/2)^{-1}\right)~,
$$
we can easily find that choosing $R_l(x)=1/(x\sqrt{f})$ is sufficient
to remove the divergence in the summation over $l$, leading to 
\begin{eqnarray}
&&\!\!\!\!\!\!
G_E(X,X')={k^2\alpha\over 8\pi^2}\sum_{n=0}^\infty 
e^{in\alpha\epsilon} \sum_{l=0}^\infty
\left[{l+1/2 \over x^3 \tilde{W}_{n}(l)} - {1\over x\sqrt{f}}\right]~.\cr
&&\label{ge3}
\end{eqnarray}
The above expression is still (ultraviolet) divergent in the coincidence
limit due to the summation over $n$. 
In a general spherically symmetric space-time, 
the expression that needs to be subtracted from
the Green function, before taking the $\epsilon\rightarrow 0$ limit, 
is known to be given by~\cite{anderson2,christensen}
\begin{widetext}
\begin{eqnarray}
\hspace{-5mm}
G_{div}&\!\!=&\!\!{k^2\over 16\pi^2}\biggl\{
{4\over \epsilon^2f} 
 +\left(\hat{m}^2+\left[\xi-{1\over 6}\right] \hat{R} \right) 
\left(
    \ln\left[
   { \hat{m}^2f\epsilon^2\over 4}\right]+2\gamma_E \right)
 - \hat{m}^2
+{1\over 12 f}\left({df\over dr}\right)^2-{1\over 6}{d^2f\over
dr^2}-{1\over 3r}{df\over dr}\biggr\}~,
\label{gfct}
\end{eqnarray}
where $\gamma_E$ is the Euler's constant. 
As shown in Ref.~\cite{anderson2}, a more suitable way to combine the
above counter-term with the original divergent expression for the Green
function is to re-express (\ref{gfct}), by using the Abel-Plana
summation formula, as 
\begin{eqnarray}
G_{div}  = {k^2\alpha\over 8\pi^2}\left[
-
\sum_{n=1}^\infty e^{in\alpha\epsilon}\left\{
{2\omega_n \over f}  
  +\left[\hat{m}^2 
    +\left(\xi-{1\over 6}\right) \hat{R}\right] {1 \over \omega_n}\right\}
+\Delta_1+\Delta_2
\right]~,
\label{gg}
\end{eqnarray}
where
\begin{equation}
\Delta_1\equiv
-\sum_{n=1}^{\infty}\left\{
{2\over f}\left(\sqrt{\omega_n^2+ \hat{m}^2f}
 -\omega_n- {\hat{m}^2 f\over 2\omega_n} \right) 
-\left(\xi-{1\over 6}\right) 
\hat{R}\left({1\over \sqrt{\omega_n^2+ \hat{m}^2f}}
-{1\over \omega_n}\right)\right\}~,
\label{Delta1}
\end{equation}
and
$\Delta_2 = \Delta_{2,1}+\Delta_{2,2}+\Delta_{2,3}$ 
with 
\begin{eqnarray*}
\Delta_{2,1}&\equiv&
{ \hat{m}^2\over 2\alpha}\ln \left( \hat{m}^2f\right)-{
 \hat{m}^2\over \alpha}\ln \left(\alpha + 
\sqrt{\alpha^2+ \hat{m}^2f}\right) + { \hat{m}^2\over 16\pi^2}\nonumber\\
&&\quad +{2i\over f}\int_{0}^\infty {dt\over e^{2\pi t}-1}\left(
\sqrt{\alpha^2(1+it)^2+ \hat{m}^2f}
-\sqrt{\alpha^2(1-it)^2+ \hat{m}^2f}
\right)~,
\end{eqnarray*}
\begin{eqnarray*}
\Delta_{2,2}&\equiv&{(\xi-1/6) \hat{R}\over 2}\left[
{1\over \sqrt{{\alpha^2}+ \hat{m}^2f}}
-{2\over \alpha}\ln\left(\alpha+ \sqrt{{\alpha^2}+ \hat{m}^2f}\right)
+{1\over \alpha}\ln \left( \hat{m}^2f\right)\right. \cr
&&\quad +\left.2 i \int_0^{\infty}{dt\over e^{2\pi t}-1}
\left(
{1\over \sqrt{{\alpha^2}(1+it)^2+ \hat{m}^2f}}
-{1\over \sqrt{{\alpha^2}(1-it)^2+ \hat{m}^2f}}
\right)\right]~,\cr
\Delta_{2,3}&\equiv&
{1\over 2\alpha}\left({1\over 12 f}f^{'2}-{1\over
 6}f''-{1\over 3 r}f'- \hat{m}^2\right)~.
\end{eqnarray*}
After subtracting (\ref{gg}) from (\ref{ge3}), we obtain 
an expression for the renormalized Green function, and its full coincidence
 limit $\epsilon \rightarrow 0$, which provides the
renormalized vacuum polarization,  
can be taken: 
\begin{eqnarray}
\langle\phi^2(X)\rangle 
=G^{(ren)}_E(X,X)&=&
{k^2\alpha\over 8\pi^2}\left\{ -\Delta_1- \Delta_2+
\sum_{l=0}^\infty\left({(l+1/2)} {1\over x^3 \tilde{W}_{0}(l)}-{1\over
		  x\sqrt{f}}\right)\right.
\nonumber\\&&
\left.
+2\sum_{n=1}^\infty \left[ \sum_{l=0}^\infty\left({(l+1/2)} {1\over x^3
					     \tilde{W}_{n}(l)}-{1\over
					     x\sqrt{f}}\right)
+{2\over f}\omega_n 
+\left( \hat{m}^2+(\xi-1/6) \hat{R}\right){1\over \omega_n} \right]\right\}~.
\label{rgf}
\end{eqnarray}
In the above relation we have separated, for convenience, the 
contribution from $n=0$. As shown in \cite{candelas}, 
this $n=0$ contribution vanishes in the
Schwarzschild case, but it does not in the asymptotically AdS case.
%
%

\section{vacuum polarization}
\label{sec6}
\subsection{Summation over $l$ and Regularity of the Green function}
We explain how we perform the summation over $l$ in (\ref{rgf}) 
in this subsection. 
At the same time, we explicitly demonstrate the 
finiteness of the summation over $n$. 
Expression (\ref{rgf}) 
should be, by construction, finite. However, as one can
immediately notice, the contribution of each term in Eq.~(\ref{rgf}) 
is not separately finite. 
We need to combine them before taking the summation so as to 
cancel with each other. 
In the following we will demonstrate the finiteness of (\ref{rgf})
explicitly. This, in principle, is not necessary, since the 
point-splitting regularization with subtraction of the appropriate 
counter-terms guarantees the finiteness of the renormalized 
Green function. However, aside from being a non-trivial check of the
calculations, 
knowing explicitly how the various divergent pieces cancel with each
other suggests a convenient strategy for the subsequent computation.

First of all, 
let us rearrange the renormalized Green function by 
adding and subtracting its WKB counterpart:
\begin{eqnarray}
\langle\phi^2(X)\rangle =
G^{(ren)}_E(X,X)
&=&
{k^2\alpha\over 8\pi^2}\left\{ -\Delta_1-\Delta_2+\Upsilon_0+
\Sigma_1+\Sigma_2\right\}~, 
\label{rgf2}
\end{eqnarray}
where
\begin{eqnarray}
\Sigma_1 &\equiv&
\sum_{l=0}^\infty{(l+1/2)} \left({1\over x^3
      \tilde{W}_{0}(l)}-{1\over x^3 {W}_{0}(l)}\right)+
2\sum_{n=1}^\infty \sum_{l=0}^\infty{(l+1/2)} \left({1\over x^3
      \tilde{W}_{n}(l)}-{1\over x^3 {W}_{n}(l)}\right)~, \qquad
\cr
\Sigma_2&\equiv&2\sum_{n=1}^\infty 
\left[\Upsilon_n
+{2\over f}\omega_n +\left( \hat{m}^2+\left(\xi-{1\over 6}\right) \hat{R}\right)
{1\over \omega_n}  \right]~,
\end{eqnarray}
and
\begin{equation}
\Upsilon_n\equiv
\sum_{l=0}^\infty \left({l+1/2\over x^3 {W}_{n}(l)}-{1\over
		   x\sqrt{f}}\right)~. 
\end{equation}
The terms other than $\Sigma_2$ are manifestly finite. 
The term $\Sigma_2$ is cumbersome to evaluate, although
straightforward. 
Once again, we can proceed in the standard way. 
Making use of the Abel-Plana summation formula, we rewrite 
$\Upsilon_n$ as
\begin{eqnarray}
\Upsilon_n
&=&
{1\over x^3}\left[{1\over 4}
  {1\over W_n(0)}
+\left(
\int_0^\infty\left({z+1/2\over W_n(z)}- {x^2\over\sqrt{f}}
\right)dz 
- {x^2\over 2\sqrt{f}}\right.\right)
+\left. i\int_0^\infty {dz\over e^{2\pi z}-1}\left({iz+1/2\over
			W_n(iz)}-{-iz+1/2\over W_n(-iz)}\right)
\right]~.
\label{ap}
\end{eqnarray}
\end{widetext}

The contribution to $\Sigma_2$ 
from the first term in the square brackets of Eq.~(\ref{ap}) is 
\begin{eqnarray}
{\cal P}_1&\equiv& {1\over 2x^3}\sum_{n=1}^\infty {1\over W_n(0)}\cr
&=&{1\over 2x^3}\sum_{n=1}^\infty \left({1\over\sqrt{\Phi(0)}}
 -{\Psi(0)\over 4\Phi^{3/2}(0)} \right)\nonumber\\
&=&{1\over 4x^2}\bigg\{
2\WW_1 + \left(
x a_1 - 2\tilde a_2 - 3a_3 \right) \WW_3 
\cr && \qquad 
-x^2 \bigg[
\left({xa_1\over 2}-2\tilde a_2
\right)\left( 2\sigma + x\sigma'\right) 
\cr &&\qquad \qquad \qquad 
     -a_3\left(3\sigma- \frac{x^2\sigma''}{2}\right)
\bigg] \WW_5
\cr && \qquad 
-{x^4\tilde a_2\over 2} \left( 2\sigma + x\sigma'\right)^2 \WW_7\bigg\}, 
\label{P1}
\end{eqnarray}
where 
we have substituted $1/W_n(z)$, given by Eq.~(\ref{dsw1}), 
in the first equality and the explicit expression of $\Psi$, 
given by Eq.~(\ref{defPsi}), in the second equality. 
Here we have introduced  
generalized $\zeta$-functions of Epstein-Hurwitz type, defined by 
\begin{equation} 
\WW_q\equiv \sum_{n=1}^\infty (\omega_n^2+x^2\sigma(x))^{-q/2}, 
\end{equation}
which occur in the computation
of functional determinants of Laplacians on Riemann spheres (Some
explicit applications can be found, for instance, in
Refs.~\cite{elizalde,brevik}.)

We refer to the $n$-th term of the divergent piece 
as ${\tt div} \left[{\cal P}\right]_n$. 
To be precise, here we define ${\tt div} \left[{\cal P}\right]_n$
by the two leading terms in the large $n$ expansion 
in the form of $b_1\, \omega_n+b_2\, \omega_n^{-1}$, 
where $b_1$ and $b_2$ are constants. 
In the above expression (\ref{P1}) 
only the term with $\WW_1$ is divergent. 
Hence, we can easily extract the divergent part of
${\cal P}_1$ as 
\begin{eqnarray}
{\tt div}\left[{\cal P}_1\right]_n ={1\over 2x^2 \omega_n}~.
\label{divp1}
\end{eqnarray}
All the sub-leading terms are finite.

The next term gives
\begin{eqnarray}
{\cal P}_2&\equiv &{2\over x^3}
 \sum_{n=1}^\infty\left[\int_0^\infty
    \left({z+1/2 \over W_n(z)} -{x^2\over \sqrt{f}}\right)dz
     -{x^2\over 2\sqrt{f}}
\right]
\cr
&=& {\cal P}_{2,1}
    +{\cal P}_{2,2},
\end{eqnarray}
where we have defined
\begin{eqnarray}
{\cal P}_{2,1} & \equiv & 
{2\over x^3}\sum_{n=1}^\infty\left[ \int_0^\infty\left({z+1/2\over \Phi^{1/2}}
-{x^2\over \sqrt{f}}\right)dz
     -{x^2\over 2\sqrt{f}}\right]~,\nonumber\\
{\cal P}_{2,2} &\equiv& 
-{1\over x^3}
\sum_{n=1}^\infty\int_0^\infty {(z+1/2) \Psi\over \Phi^{3/2}}dz~.\nonumber
\end{eqnarray}
Here the integrations over $z$ can be performed easily. 
After integration over $z$, ${\cal P}_{2,1}$ reduces to 
\begin{eqnarray}
{\cal P}_{2,1} 
=  -{2\over f}\WW_{-1}. 
\label{P21}
\end{eqnarray}
Hence, we can extract 
the divergent part of ${\cal P}_{2,1}$ as 
\begin{eqnarray}
{\tt div} \left[{\cal P}_{2,1}\right]_n 
=-{2\over f}\left(\omega_n  -{\sigma \over 2\omega_n}\right). 
\label{divI}
\end{eqnarray}
Also, we perform the integration over $z$ in ${\cal P}_{2,2}$,
substituting the explicit expression of $\Psi$, to obtain
\begin{eqnarray}
{\cal P}_{2,2}
&=&{a_1\over 6f}
\bigg\{x
\left(10 - 2 {xf'\over f}\right)\WW_1 
\cr &&\qquad \quad 
- x^3\left( 2\sigma + x\sigma ' \right)
 \WW_3
\bigg\}
\cr &&
-{\tilde a_2\over 30 f}\biggl\{
4\left( 43 - \frac{18xf'}{f} + 2\frac{x^2{f'}^2}{f^2}
  \right)\WW_1
\cr &&\qquad \quad 
 - 4x^2\left( 7 - {xf'\over f} \right) 
\left(2\sigma  + x \sigma'
\right) \WW_3 
\cr &&\qquad \quad 
+ 3x^4\left( 2\sigma + x\sigma ' \right)^2
\WW_5
\bigg\}
\cr &&
-{a_3\over 6f}\bigg\{
\left(46 - 16{xf'\over f} + 2{x^2f''\over f}
\right)\WW_1 
\cr &&\qquad \quad 
- \left(6x^2\sigma - x^4\sigma''\right) \WW_3
\biggr\}.
\label{P22}  
\end{eqnarray}
Since the divergence in the above expression is contained 
only in the terms proportional to $\WW_1$, 
we can easily extract the 
divergent part as 
\begin{eqnarray}
{\tt div} \left[{\cal P}_{2,2}\right]_n&=&
{1\over 6 \omega_n}
\left\{-{13f\over 2x^2}+{5f'\over x}
 -f''
\right\}, 
\label{divJ}
\end{eqnarray}
where we have substituted explicit forms of 
$a$'s given in Eqs.~(\ref{defa}) and (\ref{defalpha}). 

The last term to evaluate is 
\beq
{\cal P}_3={2i\over x^3}\sum_{n=1}^\infty \int_0^\infty {dz\over e^{2\pi z}-1}\left({iz+1/2\over W_n(iz)}-{-iz+1/2\over W_n(-iz)}\right)~.
\label{p3int}
\eeq
Due to the exponential fall-off, the dominant contribution to the integral comes
from the $z\sim 0$ region. Hence, we can evaluate this term by 
expanding the part enclosed by the parentheses as
\begin{equation}
\left({iz+1/2\over W_n(iz)}-{-iz+1/2\over W_n(-iz)}\right)
 =-i\sum_{j=1}^\infty c_{nj} z^{2j-1}. 
\label{defcnj}
\end{equation}
The convergence of this series will be fast for large $n$, 
while it will be slow for small $n$. Therefore we divide ${\cal P}_3$ 
into two parts as 
\beq
{\cal P}_3={\cal P}_{3,1}
+{\cal P}_{3,2}, 
\eeq
where 
\begin{eqnarray}
 {\cal P}_{3,1}
&\equiv& {2\over x^3}\sum_{n=1}^\infty \int_0^\infty 
{dz\over e^{2\pi z}-1}\sum_{j=1}^{j_{\rm max}} c_{nj} z^{2j-1}~,\cr
 {\cal P}_{3,2}
&\equiv& {2i\over x^3}\sum_{n=1}^\infty \int_0^\infty {dz\over e^{2\pi
z}-1}
\Bigg\{\bigg({iz+1/2\over W_n(iz)}
\cr &&\qquad \quad 
-{-iz+1/2\over W_n(-iz)}\bigg)
+i\sum_{j=1}^{j_{\rm max}}c_{nj} z^{2j-1}
\Bigg\}, 
\label{P32}
\end{eqnarray}
with an appropriately chosen value of $j_{\rm max}$. 
Then, on one hand, 
${\cal P}_{3,2}$ can be evaluated numerically. 
As long as $j_{\rm max}$ is sufficiently large, 
the summation over $n$ converges rapidly.  
On the other hand, 
we can perform the integration over $z$ in ${\cal P}_{3,1}$ 
analytically term by term 
using the formula $\int_0^\infty z^{2j-1}(e^{2\pi z}-1)^{-1}
dz=\Gamma(2j)\zeta_R(2j)/(2\pi)^{2j}=(-1)^{j-1}B_{2j}/4j$, where 
$\zeta_R$ is the Riemann $\zeta$-function and 
$B_1 = 1/2, B_2 = 1/6, B_3 = 0, B_4 = 1/30, B_5 = 0, B_6 = 1/42, B_7 =
0, B_8 = 1/30, \cdots $ are the Bernoulli numbers. 
Finally, we obtain 
\begin{eqnarray}
 {\cal P}_{3,1}
&\equiv& {2\over x^3}\sum_{n=1}^\infty 
 \sum_{j=1}^{j_{\rm max}} c_{nj} 
 {(-1)^{j-1}\over 4j}B_{2j}~. 
\label{P31}
\end{eqnarray}
Since $c_{nj}$ is $O(1/\omega_n^{2j-1})$ for large $n$, 
only the part with $j=1$ is divergent. 
Thus, using $c_{n1}\simeq-2/W_n(0)$ and $B_2=1/6$, we have  
\beq
{\tt div} \left[{\cal P}_3\right]_n =-{1\over 6x^2\omega_n}~.
\label{divp3}
\eeq
It is now a matter of trivial algebra to combine together (\ref{divp1}),
(\ref{divI}), (\ref{divJ}) and (\ref{divp3}), 
to show that they cancel
the contribution from the last line in (\ref{rgf}), leaving a
well-behaved expression.

The same procedure applies in computing $\Upsilon_0$, too. 
We just need to eliminate the summation over $n$ setting 
$n$ to zero in the expressions for ${\cal P}$'s. 
For the evaluation of the counter part of ${\cal P}_3$ 
in computing $\Upsilon_0$, 
we do not need to divide it into two pieces like ${\cal P}_{3,1}$ and 
${\cal P}_{3,2}$ since there 
is no infinite summation over $n$. 
  
The remaining task is composed of two parts; numerical calculation 
and analytic summation over $n$. 
The former is necessary for 
$\Delta_2$, $\Upsilon_0$, ${\cal P}_{3,2}$, 
but infinite summation over $n$ 
is unnecessary for these terms. 
Only the leading finite number of terms give sufficiently 
precise approximation.  
The other terms, i.e.  
$\Delta_1$ and the other ${\cal P}$'s, 
can be handled completely analytically. 
Moreover, from the above demonstration, we find that all the 
summations over $n$ which appear in ${\cal P}$'s 
are written in terms of ${\cal Z}_n$ and its counter-terms 
in the form $\sum_n b_1\omega_n+b_2\omega_n^{-1}$. 
$\WW_{-1}$ and $\WW_{1}$ 
contain divergences, and hence they require the counter-terms. 
The coefficients of the counter-terms, $b_1$ and $b_2$, are 
appropriately chosen so as to cancel these divergences. 
Hence, we can simply define a regularized $\WW_{-1}$ and $\WW_{1}$ 
by including the counter-terms proportional to $\sum_n \omega_n$ or $\sum_n \omega_n^{-1}$ so that they are finite. 
We denote such regularized $\WW_{-1}$ and $\WW_{1}$ as 
$\tilde \WW_{-1}$ and $\tilde\WW_{1}$, respectively. 
Then, summing up the regularized pieces gives the correctly 
renormalized value. 
%
%

\subsection{Summation over $n$}
In the previous subsections, we performed the summation over $l$
in the renormalized expression for 
$\langle \phi^2 \rangle$, Eq.~(\ref{rgf2}). 
The expression is composed of two parts; one is 
the part that requires numerical evaluation 
and the other is the part 
that requires infinite summation over $n$. 
In this subsection we will carry out the evaluation of these 
expressions. 

It is straightforward to perform the necessary numerical computations. 
The term $\Sigma_1$ requires evaluating 
$\tilde W_n$ and hence the exact mode functions, $p^n_l(x)$ and $q^n_l(x)$. 
For this term, summations over $l$ and $n$ are also necessary.  
In the practical numerical computation we need to truncate these
summations at finite $l$ and $n$.  
However, this term consists of the difference 
between $\tilde W_n$ and its WKB approximant $W_n$, 
and hence it is the remainder of the approximation in this sense. 
Since the WKB approximation becomes better for large $l$ and $n$, 
the convergence is basically fast. 
In the present approximation truncated at the next-to-leading order 
in the WKB expansion, the remaining terms cause the error 
in the estimation of $\langle\phi^2\rangle$ 
of order $O(l_{\rm max}^{-2},n_{\rm max}^{-2})$. 
As we increase the order of WKB approximation, 
the error decreases more rapidly. However, the price to pay for a faster numerical convergence is that the computation of the analytic part becomes 
more complicated. 
For small $x$, the above argument is completely correct. For larger $x$, 
however, $\Phi(l)$ is 
dominated not by $\varpi$ but by $\sigma$ in wide range of $l$ and $n$. 
When $\sigma$ is dominant, WKB expansion is not a good 
approximation at all. (WKB series does not converge in this case.) 
Therefore even if we use higher order 
WKB approximant, we need to sum the difference up to very large $l$ and $n$. 
In this case, in order to reduce the computational cost, 
we can use the interpolation of the values of difference evaluated 
only at sparse grid points of $l$ and $n$. 
${\cal P}_{3,2}$ also has a summation over $n$ 
but again it is composed of the difference between 
the expression~(\ref{defcnj}) and its truncation at $j=j_{\rm max}$. 
As in the case of $\Delta_1$, 
as we increase $j_{\rm max}$, the convergence of the summation 
over $n$ becomes faster. 
Due to this property of fast convergence,
no special trick is necessary in numerical evaluation of 
this term. 

The latter part should be treated fully analytically due 
to the presence of infinite summation. 
However, it is not so difficult because 
they are all already written in terms of 
generalized zeta-functions, 
as we have seen in the previous subsection. 
These generalized zeta-functions, 
${\cal Z}_q$, are functions of 
\begin{eqnarray*}
v^2\equiv {x^2\sigma\over \alpha^2}.  
\end{eqnarray*}
We need to evaluate them as a function of $v^2$ 
only once at the beginning of the whole calculation. 
Hence, the direct sum is one possible way to compute them. 
However, there are slightly cleverer 
ways to reduce the computational cost, which we will explain below. 

One way to truncate this summation
is as follows. Formally we write the summation defining 
${\tilde {\cal Z}}_q$ as $\sum_{n=1}^\infty F_q(n)$. 
This summation can be directly performed numerically 
up to a certain value $n_*$. For the remaining part of summation, 
we simply expand $F_q(n)$ 
in terms of $v$ as 
\begin{eqnarray}
 F_q(n)=(\alpha v)^{-q} \sum_{k=1} d_{qk}\left({v\over n}\right)^{2k+1},  
\label{expansionFq}
\end{eqnarray}
where the coefficients $d_{qk}$ are independent of 
$\alpha, v$ and $n$. 
Then, we have
\begin{eqnarray}
\sum_{n=1}^\infty F_q(n)
&\approx &\sum_{n=1}^{n_*} F_q(n)\cr
 &&~  +\alpha^{-q} \sum_{k=1}^{k_{\rm max}} d_{qk}\tilde \zeta(2k+1) 
v^{2k-q+1},  
\end{eqnarray} 
where 
\begin{eqnarray*}
 \tilde \zeta(s)\equiv \sum_{n=n_{\rm max}+1}^\infty {1\over n^s}
           =\zeta(s)- \sum_{n=1}^{n_{\rm max}} {1\over n^s}~.
\end{eqnarray*}
In practice, we need to truncate the summation over $k$ at a 
certain value $k_{\rm max}$, 
which is not necessarily so large as long as $n_*$ is 
sufficiently large. In order that 
the expansion (\ref{expansionFq}) converges 
for $^\forall n\ge n_*$, $n_*^2$ must be larger than $|v^2|$. 
Hence, this method works efficiently unless $|v^2|$ is very large.

On the other hand, for a very large positive value of $v^2$, 
we can make use of a simplified version of 
the Chowla-Selberg formula 
(see for example \cite{cs,elizalde}), 
\begin{eqnarray}
S(s,\rho)&\equiv&
\sum_{n=1}^\infty (n^2+\rho^2)^{-s}\cr
& =&
 -{\rho^{-2s}\over
2}+{\sqrt{\pi}\over 2}{\Gamma(s-1/2)\over \Gamma(s)}\rho^{1-2s}
\label{chse}
\\ &&
+
{2\pi^s\over \Gamma(s)}\rho^{-s+1/2}\sum_{p=1}^\infty
p^{s-1/2}K_{s-1/2}\left(2\pi p \rho\right)~.
\nonumber
\end{eqnarray}
The above identity basically rearranges the original sum in a convenient
way, as the summation on the R.H.S. of the above equation converges very
rapidly due to the exponential fall-off of the modified 
Bessel functions 
unless $\rho$ is extremely small.  

Among $\tilde \WW_q$, $\tilde\WW_{-1}$ 
requires the most delicate treatment. 
By definition, 
\beq
\tilde\WW_{-1}=\alpha \sum_{n=1}^\infty\left(
  \sqrt{n^2+v^2}-n-{v^2\over 2n}\right).  
\label{zetamenouno}
\eeq
The Chowla-Selberg formula does not directly apply to this expression 
since the R.H.S. of (\ref{chse}) diverges for $s=1/2$. 
Of course, it is due to this divergence why we subtracted 
counter terms. The trick is to regularize the above expression as 
\begin{eqnarray*}
\tilde \WW^{\rm reg}_{-1}
\!\!\!&=&\!\!
\alpha \sum_{n=1}^\infty\left(
  \left(n^2+v^2\right)^{{1\over 2}-\epsilon}
\!\!\!-n^{1-2\epsilon}
\!\!\!-
\left({1\over 2}-\epsilon\right){v^2}
n^{-1-2\epsilon}\right)\nonumber\\
\!\!\!&=&\!\!
S(-1/2+\epsilon,v^2)-\zeta_R(-1+2\epsilon)
\cr&&\qquad \qquad  \qquad 
   -\left({1\over 2}
-\epsilon\right){v^2}\zeta_R(1+2\epsilon). 
\end{eqnarray*}
This expression reduces to $\tilde\WW_{-1}$ if we 
set $\epsilon$ to $0$, and it 
is finite for any non-negative $\epsilon$. 
Hence, the limit $\epsilon\to 0$ gives $\tilde\WW_{-1}$. 
We can use the Chowla-Selberg formula, take the limit, and obtain an explicitly regular expression, 
\begin{eqnarray}
\tilde\WW_{-1}
\!\!\!&= &\!\!\!\alpha\bigg(
{1\over 12}-{v\over 2} 
+v^2{1-2(\ln (v/2)+\gamma_E)\over 
 4}\cr &&\qquad\qquad\qquad
-{v\over \pi}\sum_{p=1}^{\infty}{K_{-1}(2\pi p v)\over p}
\bigg). 
\end{eqnarray}
Rearranging the generalized zeta functions by means of the
Chowla-Selberg formula, {\it i.e.} as a series of modified Bessel
functions,  proves to be efficient when the argument of the modified
Bessel functions is real and not small. 

For $\tilde\WW_{1}$, the regularized expression is 
simply given by 
\begin{eqnarray*}
\tilde \WW^{\rm reg}_{1}
=S(1/2+\epsilon,v)-\zeta_R(1+2\epsilon).
\end{eqnarray*}
Taking the limit $\epsilon\to 0$, we obtain 
\begin{eqnarray}
\tilde\WW_{1}
\!\!\!&= &\!\!\!-\alpha^{-1}\bigg(
\frac{1}{2 v} + \ln (v/2) + \gamma_E 
-2\sum_{p=1}^{\infty}{K_{0}(2\pi p v)}
\bigg). 
\end{eqnarray}
The other generalized zeta functions $\WW_q$ with 
$q>1$, which do not require further regularization, 
are evaluated directly applying 
the Chowla-Selberg formula. 
We find 
\begin{eqnarray}
\WW_q&=& \alpha^{-q}\bigg(
-{v^{-q}\over 2}+{\sqrt{\pi}\Gamma((q-1)/ 2)\over 2
\Gamma(q/2)}v^{1-q}
\cr &&
+
{2\pi^{q/2}\over \Gamma(q/2)}v^{{1-q\over 2}}\sum_{p=1}^\infty
p^{{q-1\over 2}}K_{{q-1\over 2}}
 \left(2\pi p v\right)\bigg)~.
\end{eqnarray}

Finally, we mention how to evaluate $\Delta_1$. 
It can be evaluated in an analogous manner.  
The first and the second terms in $\Delta_1$ have 
exactly the same forms as $\tilde\WW_{-1}$ and $\tilde\WW_{1}$, 
apart from the replacement of $v^2$ with $\hat{m}^2 f/\alpha^2$.  
We write it here more explicitly as 
\begin{equation}
\Delta_1=\left[
-{2\alpha \over f}\tilde\WW_{-1}
+\left(\xi-{1\over 6}\right) 
\hat{R}\tilde\WW_{1}\right]_{v^2\to \hat{m}^2 f/\alpha^2}. 
\end{equation}

\section{results}
So far, we have not specified the form of the metric. For the numerical evaluation, however, we need to fix the function $f(r)$. To illustrate the results we will choose the case of a Sch-AdS black hole, with metric given by (\ref{metric}) and $f$ by (\ref{schads}).
For notational convenience we define the quantity $\ell_{bh}\equiv2M m_p^{-2}k$, which characterizes the size of the black hole. 

The numerical analysis does not present particular difficulties and can be performed in a straightforward manner. The WKB part is expressed in terms of analytic functions and can thus be evaluated very easily. Slightly more involved is the computation of the sums. The WKB approximation to next-to-leading order ensures that the convergence goes relatively fast, as $O(l^{-5},n^{-5})$. Therefore, after obtaining the solutions for the modes numerically, the summations can be computed directly. 

\begin{figure*}[th]
\scalebox{1} {\includegraphics{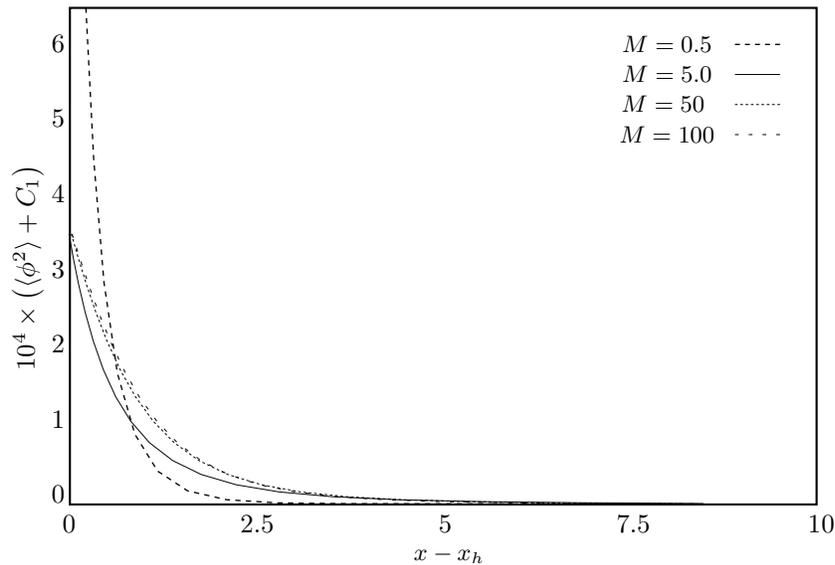}}
\caption{The figure illustrates the behavior of $\langle \phi^2 \rangle$ for conformally coupled ($\xi=1/6$ and $m=0$) fields. The curves refer to: $M=100,~50,~5,~0.5$.}
\label{figura}
\end{figure*}

After performing standard numerical checks, the various terms
contributing to $\langle \phi^2 \rangle$ can be combined and some
illustrative curves are reported for some representative values in the
case of  conformally coupled scalars ($\xi=1/6$ and $m=0$) in
Fig.~\ref{figura}. (For  $n=l=0$ and very small values of $M$, the
function $\Phi$ may become negative. In this case, the $n=l=0$ mode
has to be treated separately in the single sum $\Upsilon_0$). 
\vskip 0.4cm
Due to the complexity of the calculation, it is instructive to look at
the behavior of the vacuum polarization far away from the black hole,
where we can expect to reproduce the leading AdS space result, which can
be calculated independently. In fact, in pure AdS space the Green
function is given by 
\beq
G(X,X') = {2\over 3\pi^2}\left[3-\cosh\left(k\sqrt{d(X,X')}\right)\right]^{-1}~,
\nonumber
\eeq
where $d(X,X')$ is the geodesic distance between $X$ and $X'$. By
subtracting, as in the black hole case, the counter-terms and then
taking the coincidence limit, one gets:
\beq
\langle \phi^2 \rangle_{\small AdS} \simeq -{1\over 48\pi^2}~.
\eeq
Our results for 
Sch-AdS case can be fitted in the asymptotic limit as  
\beq
\langle \phi^2 \rangle \simeq C_1 + C_2/f~.
\eeq
with $C_1=-{1\over 48\pi^2}$. 
The first term represents the leading
contribution to the vacuum polarization. 
We have checked the universality of $C_1$ to high accuracy. 

The coefficient $C_2$ of the second term 
is hard to anticipate before calculation. 
If we use the result for the {\it vacuum polarization} 
at a finite temperature in Minkowski space,   
setting the temperature to the local Hawking temperature $k\alpha/2\pi\sqrt{f}$, 
we obtain the estimate $C_2={\alpha^2\over 48}$. 
However, our numerical result for $C_2$ is much smaller than this value 
in general. 
Since the temperature at large distance from the
black hole is very low, a typical 
energy scale for the excitation is much below the inverse curvature length. Thus, it is not surprising that the rough estimate based on  
the result in Minkowski space does not hold in the present case.

\section{discussions and conclusions}

In this paper we developed a method to compute the renormalized
expectation value of a quantum scalar field, with mass $m$ and coupling
to the curvature $\xi$, in a thermal state on a spherically symmetric,
asymptotically AdS black hole geometry. 

We followed the approach of Refs.~\cite{candelas,anderson2}, and
employed the analytic WKB approximation and the point-splitting
regularization to construct a regular expression for the coincidence
limit of the Green function. We explicitly demonstrated the regularity
of the Green function, and this allowed us to perform the
renormalization term by term.

Analogously to the asymptotically flat case, the WKB approximation
arranges the vacuum polarization in a `WKB-part' plus a 
remainder. 
One
term depends on the WKB approximants and it can be evaluated
analytically, although the computations become more cumbersome as the
order of the approximation increases. 

It is very interesting to notice that the WKB approximation organizes
the analytical part of the vacuum polarization as a series of analytic
functions. These functions take the form of generalized zeta functions,
that occur in the computation of functional determinants of Laplacians
on Riemann spheres. The coefficients of the expansions can be calculated
order by order in the WKB expansion. We explicitly showed this to
next-to-leading order in the WKB approximations. The other term is a
remainder 
 of the WKB approximation, in the sense that it depends on the
difference  between the exact solutions for the modes and their WKB
counterpart. 

In the end, both terms have to be handled numerically. The 
generalized zeta functions can be easily evaluated using 
direct summation when their argument is small. When the 
argument is not small, it is convenient to rearrange these 
functions by means 
of a simplified version of the Chowla-Selberg formula \cite{elizalde}. 
The rearranged expression contains an infinite series of modified Bessel 
functions, but it converges exponentially fast when the argument is 
not small. 
The 
`remainder' 
can be evaluated by directly solving the mode equation
numerically and taking the difference with the WKB counterparts. These
terms converge very rapidly, as we expected from the fact that the WKB
approximation works very well: at next-to-leading order the remainder is
$O(l^{-5},n^{-5})$. 

In the last part, we  explicitly illustrated the method by computing the
vacuum polarization for conformal fields on a Sch-AdS black hole
background. We finally discussed the asymptotic behavior of the vacuum
polarization and compared 
it 
with the pure AdS result which can be calculated independently. We find that our result reproduces the leading universal (constant) behavior to high accuracy. 

We are now using the methodology presented in this paper to compute the
energy momentum tensor for a quantum field on an 
asymptotically AdS black hole geometry and we hope to report on this soon.

\acknowledgements
We wish to thank Y. Sendouda for useful discussions related to the numerical computations presented in this article. The support of the 21st century COE ``Center for Diversity and Universality in Physics'' at Kyoto University, from the Ministry of Education, Sports, Science, and Technology of Japan is kindly acknowledged. AF acknowledges the support of JSPS under grant N. 19GS0219. TT is supported by Monbukagakusho Grant-in-Aid for Scientific Research Nos. 17340075 and 19540285.

\end{document}